\documentstyle[11pt,fleqn,leqno,epsfig]{article}
\begin{document}

\title { A possible Universal Origin of Hadronic Cosmic Rays  \\ 
from Ultrarelativistic Ejecta of Bipolar Supernovae }

 \author{R.Plaga\thanks
        {Telephone 49-89-32354-224;
         FAX: 49-89-3226704; email:plaga@mppmu.mpg.de}
     \\Max-Planck-Institut f\"ur Physik (Werner-Heisenberg-Institut)
     \\ F\"ohringer Ring 6
     \\D-80805 M\"unchen, Germany}

  

   \date{ Received 27 September 2001; received in revised form 3 April 2002; 
   accepted 3 April 2002 \\ Communicated by R.P.Kirshner}
     
\maketitle
\markboth{A universal Origin of Cosmic Rays}
{}

\begin{abstract}
\noindent
Based on the ``cannonball model'' for gamma-ray
bursts of  Dar and De R\'ujula
it is proposed that masses of baryonic plasma (``cannonballs''),
ejected in bipolar supernova explosions in our Galaxy
are the sources of hadronic Galactic cosmic rays (CRs) at all energies.
The propagation of the cannonballs in the Galactic disk and halo
is studied. Two mechanisms for the acceleration of the observed CRs
are proposed.
The first is based on ultrarelativistic shocks in the 
interstellar medium and could accelerate the bulk of CRs up
to the ``knee'' energy of 4 $\times$ 10$^{15}$ eV.
The second operates with second-order Fermi acceleration
within the cannonball. If the total initial 
energy of the ejected plasmoids
in a SN explosion is 10$^{53}$ ergs or higher
this second mechanism may explain the CR spectrum above the knee
up to the highest observed energies.
It is shown that together with plausible assumptions about
CR propagation in the Galactic confinement volume the observed
spectral indices of the CR spectrum can be theoretically
understood to first order.
The model allows a natural understanding of various
basic CR observations like the absence of the Greisen-Zatsepin
cutoff, the anisotropy of arrival directions as function of energy and
the small Galactocentric gradient of the CR density.
\end{abstract}
%
\section{Introduction}

\subsection{Motivation for a ``unified'' 
(single-source class) explanation
of the spectrum of Galactic cosmic rays}
\label{uni.i}
The flux of the dominating 
hadronic component of the local non-solar cosmic rays (CR),
has been measured between about 0.1 GeV and
3 $\cdot$ 10$^{20}$ eV.
Neglecting solar modulation effects,
its energy spectrum
can be well described 
by a single power-law 
that changes its slope slightly at only
two energies - at $\approx$ 4 $\cdot$ 10$^{15}$ eV
(the ``knee'') and $\approx$ 3 $\cdot$ 10$^{18}$ eV (the ``ankle'')\footnote{
Possible further small deviations are discussed in section \ref{secondknee}}.
This striking simplicity and unity of the data over 
more than 10 decades in energy originally led
most authors to ascribe 
the origin of hadronic cosmic rays to a single ``universal'' source class.
\\
In an influential review, 
Ginzburg\cite{ginz} suggested that the remnants of supernova explosions in our
Galaxy (SNRs) are the dominant source of CRs at all energies.
However, SNRs cannot
explain the origin of CRs with energies above
above
about 10$^{18}$ eV because at these energies 
charged particles are certainly
no longer confinable in the relatively
weak magnetic fields of SNRs. In another classical paper
Burbidge\cite{burb} proposed the radical
alternative of an extragalactical origin of CRs at all
energies.
In this scenario it proved difficult to understand the origin of
CRs with low energies: no extragalactic source class seems
able to fill up the universe with
the rather high total energy-density of the locally observed CRs
($\approx$ 0.5 eV/cm$^3$) within a Hubble time.
\\ 
As a compromise, an ``eclectic scenario'' for CR origin
was proposed by Morrison\cite{morrison}. 
This idea 
became widely accepted in the 1970s in the following form: 
{\it SNRs accelerate the CRs below the knee
and extragalactic sources are the source of CRs above the ankle.}
Between the knee and ankle
another class of objects - of an as yet unclear nature - is supposed
to accelerate the CRs\footnote{Biermann's\cite{bier95}  theory of CR origin 
does not make this assumption and is further discussed 
in section \ref{aimplan}}.
However, the ankle lies just at the energy where CRs become unconfined
from the Galaxy. This fact can only be causally understood
if the same universal source class supplies the cosmic rays below and above
the ankle\cite{tkac,plaga98}.
The eclectic scenario was born of distress,
namely of the difficulty that no ``universal'' source
class was known that can accelerate particles to the
highest observed energies and at the same time supply the 
rather high local energy density of CRs.
The aim of this paper is to revisit the idea
of a universal origin of Galactic CRs and 
to propose a source class capable of accelerating
nuclei to the observed energy spectrum. 

\begin{figure}[ht]
\vspace{0cm}
\hspace{0cm} \epsfxsize=9.1cm 
\epsfbox{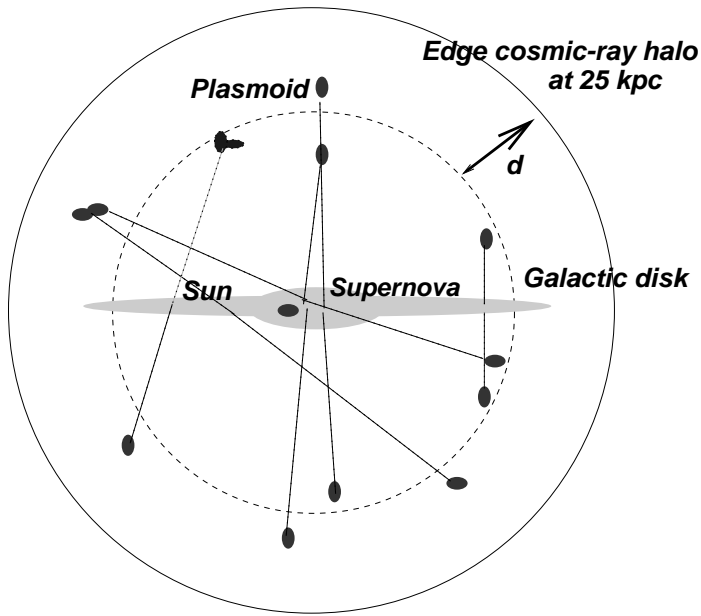}
\vspace{0cm}
\caption{A scheme of the scenario proposed here.
Energetic core-collapse supernovae in the Galactic disk and centre
eject plasmoids (``cannonballs'') 
that are slowed down to non-relativistic speeds
in the Galactic halo. These cannonballs 
are proposed to accelerate the major part of hadronic CRs at all energies.
}
\label{fig0}
\end{figure}

\subsection{Clues to the nature of the ``single source'' class
from experimental evidence on cosmic rays}
\label{clue.i}
In all probability
the universal source class resides within or
near to our Galaxy.
It is all but impossible to supply the high ambient Galactic
cosmic-ray energy density with any extragalactic source class.
The absence of the Greisen cutoff in the experimentally determined
cosmic-ray spectrum
is another strong argument
in favour of a Galactic source class.
The observed lack of an overall anisotropy of arrival directions
in UHE cosmic rays,
the very small anisotropy at all smaller energies, and
the - at best - very weak concentration of the cosmic-ray density
towards the Galactic centre forces us to conclude that the
``universal source class'' (if it exists)
has a very different spatial 
distribution than SNRs which populate the disk and central
region of the Galaxy. It seems to reside in the Galactic 
halo. 
A possible origin
of CRs at all energies
in the Galactic halo, due to Fermi acceleration in 
hydromagnetic turbulence, has been discussed
by Burbidge\cite{burb55}. Biermann \& Davis\cite{bier58} and 
Plaga\cite{plaga98}
discuss mechanisms by which a large Galactic halo may be filled
homogeneously with CRs.

\subsection{Ejecta from bipolar supernovae as sources
of Galactic cosmic rays}
\label{sn.i}
There has been recently strong 
observational evidence that core-collapse
supernovae (SNe) can produce ``long'' $\gamma$-ray bursts (GRBs)
\cite{sngrb}.
A related recent revolution in the understanding of SNe is observational
evidence that apparently
all core-collapse SNe are asymmetric ``bipolar''
explosions\cite{bipolar}.
Observations cannot yet directly decide yet if: 
\\
a. only very few ``exceptional'' 
SNe emit GRBs\cite{pac97} or
\\ 
b. a sizeable fraction of or even all
core-collapse SNe emit GRBs jet like {\it into a narrow cone
along a symmetry axis}\cite{cen}. 
\\
There is direct observational evidence that 
GRBs are emitted by matter at ultrarelativistic speeds.
Circumstantial evidence suggests that GRBs involve
rather strong magnetic fields. This suggests that
they are excellent candidates for the acceleration of
CRs at all energies.
Especially if option b. above should be correct, 
the following question will be decisive for CR physics:
\\
{\bf Do the well known non-relativistic
isotropic ejecta
(SNRs) or the only recently
discovered relativistic jet-like ejecta 
dominate the CR production of all supernovae?}
\\
It has been suggested that
the very high ejection speeds in conjunction
with strong magnetic fields allow acceleration to much higher
energies than SNRs\cite{grb2,grb3}. This makes jet-like SN ejecta
serious {\it candidates} for the long sought universal source class of
hadronic CRs at {\bf all} energies\cite{dar98b,darplaga}.
\\
The potential importance of Gamma-Ray bursters for the origin
of Galactic CRs at energies above the knee had been first studied 
by Milgrom and Usov\cite{grb4} and Dar\cite{dar98b}. Recently
Pugliese et al.\cite{pugliese2} concluded that GRBs are 
unimportant as sources
of CRs at all energies whereas Dermer\cite{dermer2000} argues that objects
related to GRBs might
supply most of the observed low-energy hadronic CRs.
The reason for this disagreement is the use of very different GRB
models.

\subsection{Aims and plan of this paper}
\label{aimplan}
The basic idea of the present paper - 
that ultrarelativistic ejecta from stellar-collapse events accelerate
the dominant part of
hadronic CRs at {\it all} energies and release it in the Galactic halo 
(fig. \ref{fig0}) -
was first proposed by Dar \& Plaga\cite{darplaga}.
At that time the nature of the GRB emitting collapse-events was unclear 
(supernovae were only mentioned as one possibility) and
they were called ``Galactic Gamma-Ray Bursters (GGRBs)''.
In the present paper the nature of these collapse events, the propagation
of their ejecta and the CR-acceleration mechanisms operating near them
are discussed more quantitatively.
\\
Section \ref{uni.l}
discusses what one can conclude
about the properties of a hypothetical ``single-source class'' 
purely from the phenomenology of CRs.
I conclude
that universal sources must be mainly
located outside the solar circle, in the Galactic halo.
\\
Further observational evidence in favour of such a location of
the CR sources is discussed in
section \ref{clue.l}. 
The Galacto-centric distribution
of $\gamma$-rays 
and anisotropy of charged CRs support the idea that
CRs below the ankle have their origin mainly (but not
exclusively) at Galacto-centric distances exceeding the solar one. 
\\
A candidate for the ``single-source class'' is proposed in
section \ref{sn.l}:
ultrarelativistic plasmoids
ejected in bipolar supernova explosions (``cannonballs''
\cite{darrujula}). The motion of cannonballs in the Galactic
disk and halo is discussed.
Section \ref{accel}
studies mechanisms with which
these ejecta produce the hadronic CR spectrum at all energies (\ref{fig4}).
\\
This scenario is related to ideas of Bierman\cite{bier95} who proposes
an origin of Galactic CRs up to the ankle in SNRs and an origin
of the higher energy CRs in extragalactic jets.
My work can be seen as a proposal to ``replace'' 
the non-relativistic SN remnant
with ultrarelativistic supernova ejecta. These ejecta (``cannonballs'') 
``assume'' the role of an extragalactic jet
in a later stage of their evolution. 
\\
The ideas of sections \ref{uni.l} and \ref{sn.l}
are logically independent. It is conceivable that there is
a single-source class, but I err in its identification, 
or that cannonballs contribute to the CR spectrum only in
certain limited energy ranges.
Section \ref{ofacts} confronts expectations in a scenario
where cannonballs dominate the CR production at all energies with
observations.
Section \ref{sum} summarises the main assumptions, achievements
and predictions of this paper.
This paper only treats hadronic CRs, 
leaving aside the origin of electrons
in the Galactic CRs as a completely separate issue.

\section{Properties of a hypothetical
single-source class of cosmic rays derived from observations}
\label{uni.l}
In this section I try to derive the likely properties of a 
hypothetical ``universal'' CR-source class from observed
CR properties without making assumptions about the physical
nature of the sources. In this I optimistically take it for
granted that a ``simple'' understanding of the CR spectrum -
e.g. without accidental cancellations of additional effects - is
possible.
\subsection{Energy dependence of 
cosmic-ray diffusion coefficient assumed in this work}
\label{diffcoe}
In this work I will assume that the cosmic-ray propagation
is purely diffusive. The dependence of the diffusion
coefficient D on the total energy per nucleus E is usually parametrised as:
\begin{equation}
D \sim  E^{\alpha} 
\label{diff1}
\end{equation}
There are two theoretically
motivated values for $\alpha$. A plasma with turbulent inhomogeneities  
following a ``Kolmogoroff
spectrum'' would lead to a diffusive CR motion 
with $\alpha$=0.33, alternatively, with a `` Kraichnan spectrum''  
$\alpha$=0.5 would apply\cite{cesarsky}.
Direct determinations of the electron-density fluctuations in the interstellar
medium\cite{armstrong} indicate a turbulence spectrum
that leads to a 
power-law dependence of D
according to eq.(\ref{diff1}) up to very high energies\footnote{up to scales
corresponding to the Larmor radius
of $\approx$ 10$^{17}$ eV particles in the Galactic magnetic field.
}.
However, such measurements are not precise 
enough to determine $\alpha$.
\\
CR data obtained at low energies indicate
$\alpha$ $\approx$ 0.6-0.7\cite{bere}.
This value can be made compatible with
both theoretical values by postulating varying degrees of ``re-acceleration''
during the propagation of cosmic rays. There is recent experimental
evidence for some re-acceleration\cite{connell} but 
it is not yet clear which value for $\alpha$ is preferred 
by these data.
\\
It is generally accepted and plausible that CRs are magnetically
confined to the Galaxy up to energies of about E$_{\rm ankle}$ $\approx$
3 $\times$ 10$^{18}$ eV\cite{bere}.
At lower energies
the power-law index of the CR spectrum 
changes only once by 0.3 at the knee.
In principle the ``knee'' could signal a change from
$\alpha$=0.33 to $\alpha$=0.5. However, this is unlikely 
because the observed change in the power-law index of $\approx$
0.3 is 
significantly larger than the difference between the
theoretically preferred indices of (0.5 - 0.33). I conclude
that in any simple scenario - where $\alpha$ has one of the
theoretically preferred values at all energies -
$\alpha$ is constant at either 0.33 or
0.5 at {\it all} energies up to the ankle.
In the next section I will argue that 
- at least in a scenario with
``universal sources'' - $\alpha$=0.5 is the preferred value.

\subsubsection{An argument in favour of $\alpha$=0.5 at all energies
valid under the assumption that universal sources of CRs exist
(can be omitted on first reading)}
\label{propa} 
Let d be the typical distance of the universal sources from
the edge of the Galactic confinement volume (see fig.(\ref{fig0})).
The CR flux at an energy of

E$_1$ = 40 GeV is a factor K=(E$_1$/E$_{\rm ankle}$)$^{-\alpha}$
higher than at energy E$_{\rm ankle}$ due to a K times longer
confinement time $\tau$.
The confinement time at energy E$_1$
is given as $\tau_1$ $\approx$  d$^2$/D$_1$, where D$_1$
is the diffusion coefficient at energy E$_1$\footnote{
The diffusion coefficients are specified by Strong \& Moskalenko
(1998) at a rigidity of 5 GV.  For intermediate masses (A=16)
this corresponds to an energy per nucleus E$_1$ $\approx$ 40 GeV. }.
For a halo size h $\approx$ 20 kpc (a large halo size is preferred
for for the universal-source scenario, see section \ref{hlower})
D was determined from measured isotope-ratios 
as approximately
9 $\times$ 10$^{28}$ cm$^2$/sec\cite{strong1}.
The confinement time at energy E$_{\rm ankle}$ is approximately
$\tau_a$ $\approx$ d/c, the value for unconfined motion.
Setting K=$\tau_1$/$\tau_a$ I obtain:
\begin{eqnarray}
d \approx (E_1/E_{\rm ankle})^{- \alpha} \times D_1/c  
\nonumber
\\
\approx {\rm 8 \, kpc \, for} \, \alpha=0.5
\nonumber
\\
\approx {\rm 0.3 \, kpc \, for} \, \alpha=0.33
\end{eqnarray}
The confinement volume is certainly extended by more than
10 kpc in our scenario (see section \ref{hlower}) and therefore
the case $\alpha$=0.33 would lead to a ``crowding'' of the universal
sources in the outer $<$ 3 $\%$ of the confinement volume.
This seems rather unnatural and we therefore assume in the
following that {\it $\alpha$ = 0.5 at all energies $<$ E$_{\rm a}$}.
\\
To summarise, the assumption of a E$^{0.5}$
dependence of CR diffusion at all energies where Galactic
confinement occurs is theoretically motivated, in agreement
with all data and allows a natural spatial distribution of universal
sources.

\subsection{The value of the spectral indices of the source
spectrum, before Galactic modulation(\cite{darplaga})}
\label{sourcespec}
The assumption of $\alpha$=0.5 (eq.\ref{diff1}) fixes
the spectrum of the ``universal CR sources''. 
Let the source spectrum be given as:
\begin{equation}
F_{\rm source} \sim E^{\delta}
\label{source2}
\end{equation}
If this source emits into a volume that diffusively confines 
CRs the observed spectrum within this volume is given as:
\begin{equation}
F_{\rm observed}(E<{E_{\rm ankle}}) \sim E^{(\delta+\alpha)} \sim E^{\gamma_2}
\end{equation}
This spectrum holds
at all energies where cosmic rays are magnetically confined
to the Galaxy (i.e. up to the ankle).
At higher energies the pure source spectrum is expected again:
\begin{equation}
F_{\rm observed}(E>{E_{\rm ankle}})
 \sim E^{\delta} \sim  E^{\gamma_3}
\end{equation}
at higher energies.
The observed index $\gamma_2$ between knee and ankle
is (${\delta_2+\alpha}$)=$-$ 3. I infer for the source
index $\delta_2$=$-$2.5, a value that is compatible within 1.3 $\sigma$
with world-mean of experimental determinations of the 
index above the ankle of $\gamma_3$=$-$2.75 $\pm$ 0.2\cite{watsonagano}
as it must be within my model.
According to section \ref{diffcoe}
I assume no breaks in the energy dependence of the diffusion
coefficient which governs CR propagation, so the
``knee'' in the CR spectrum must be a source feature.
The spectral index of the source must then decrease by 0.3
at the knee, so the $\delta_1$=$-$2.2 at energies below the knee.
This general scenario for the origin of the total 
CR spectrum is illustrated in fig.(\ref{figuspec}) (taken from
Ref.\cite{darplaga}).
\begin{figure}[ht]
\vspace{0cm}
\hspace{0cm} \epsfxsize=9.1cm 
\epsfbox{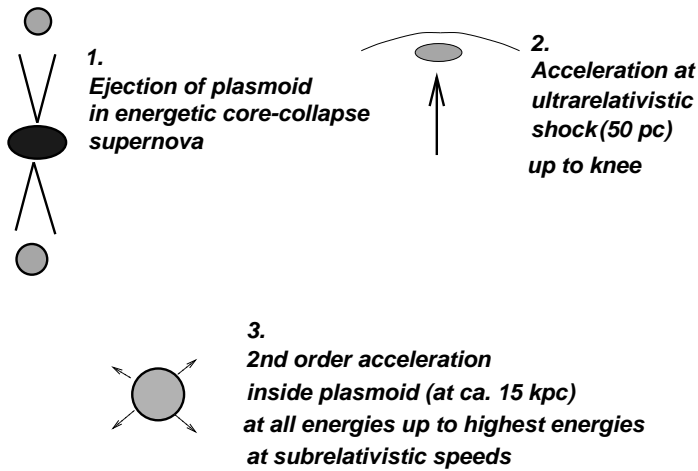}
\vspace{0cm}
\caption{A schematic sketch of the main steps in the
scenario discussed here. The shaded balls symbolise the cannonballs.
The thin un-arrowed line in the second image stands for a shock wave
driven by the cannonball into the interstellar medium.
Step 1 is further explained in
section \ref{cb}, step 2 in section \ref{1st} and step 3 in section \ref{syro}.
}
\label{fig4}
\end{figure}

\subsubsection{Unsettled issues: structure at the ``knee'' and
the ``second knee''}
\label{secondknee}
The discussion of the spectrum in the previous section is
oversimplified. Erlykin and Wolfendale\cite{erlykin2000} have argued 
that an additional component that raises the flux by about a
factor 2 in a narrow energy interval
near the knee at 4 $\times$ 10$^{15}$ eV is indicated by the data.
Detailed models for the ``universal sources'' have to deal
with this feature. 
\\
At energies around 4 $\times$ 10$^{17}$ eV the spectrum 
steepens by about 0.2 in the power law index, thus forming
a ``second knee''\cite{watsonagano}.
In our scenario it seems natural that the transition between diffusion in
a turbulent plasma and approximate straight-line motion
is gradual and therefore a steepening occurs before the
ankle.
\\
I defer more detailed 
speculations about the nature of
these finer details of the spectrum to a later time.


\subsection{Spatial location of the universal CR sources}
\label{ghalo}
\subsubsection{Upper limit on distance of universal sources
from Galactic centre: isotopic data of low-energy CRs}
\label{hupper}
The data on isotopic composition of CRs yield an upper limit on
the size of the Galactic halo in which 
magnetic confinement takes place. Because the CR spectrum
can only arise in the way outlined in section \ref{sourcespec} if the
particles are initially magnetically confined, the maximal
size of the halo is also an upper limit on the distance
of the universal CR sources from the Galactic centre.
\\
The size h of a cosmic-ray halo around the Galaxy is only
weakly constrained by results on the abundance of unstable 
isotopes in low-energy cosmic rays\cite{strong1}. 
A halo with h $>$ 4 kpc is required to exist, and h $\gg$ 20 kpc
seems to be ruled out with all parameter choices, so we assume
a maximal allowed size of 30 kpc\footnote{
This upper limit is quite uncertain mainly for four reasons.
1. Strong and Moskalenko\cite{strong1} 
do not take into account uncertainties of the
input parameters, which are considerable. 2. The spatial distribution of 
universal CR sources is expected to be different from the one assumed
by Strong and Moskalenko. 3. CR-electron sources might have 
a very different spatial distribution (following the
SNR distribution) 
from the one of CR-hadron sources
(following the cannonball distribution in our scenario).
4. The plausible possibility of a diffusion coefficient 
that is smaller in the Galactic disk than the halo\cite{ginz76} 
is not considered. This possibility gains in potential importance
if sources are not located in the Galactic disk.}.

\subsubsection{Lower limit on distance of universal sources
from Galactic centre: isotropy of UHE CRs}
\label{hlower}
No significant anisotropies have been found in the 
sky distribution of CRs with energies above
4 $\times$ 10$^{19}$ eV\cite{watsonagano,wibig}.
However,
with a world statistics of 114 events, anisotropies 
on the order of up to 30 $\%$ amplitude cannot be ruled out.
At these energies - far above the ``ankle'' -
light nuclei are expected to move in
straight lines to good approximation in our scenario.
Also light nuclei comprise a major fraction of all CRs in ``universal scenario''.
(fig. \ref{chem1}). We conclude that a major part ($>$ 50 $\%$)
of all universal sources must lie at distances beyond the solar circle
(i.e. Galacto-centric distances $>$ 10 kpc).

\subsubsection{The likely location of ``universal'' CR sources}
The above consideration indicate a typical location of the
``universal sources'' at about 10 - 30 kpc from the Galactic centre.
For the reasons mentioned above these limits (especially
the upper one) are quite
uncertain. A detailed
simulation using the code of Strong $\&$ Moskalenko\cite{strong1}
and taking into account the special features of the universal source
class is urgently required.
\begin{figure}[ht]
\vspace{0cm}
\hspace{0cm}\epsfxsize=10.1cm \epsfbox{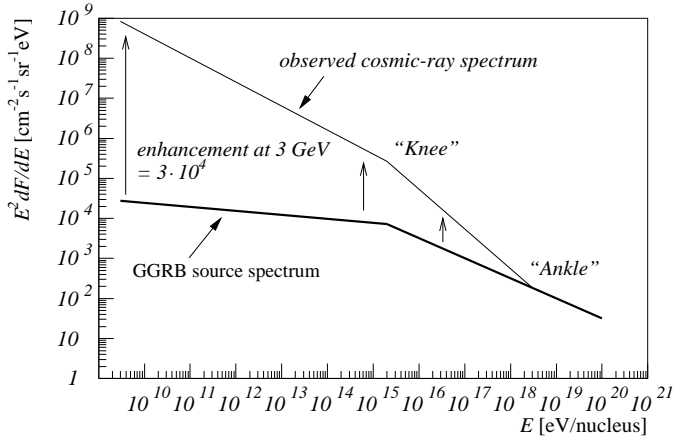}
\vspace{0cm}
\caption{ The observed flux of cosmic rays
(thin line) as a function of primary energy E 
is well described by a power law that changes
its slope sharply at only two
energies, the ``knee'' and the ``ankle''.
At energies below the ankle
it is enhanced (by a factor (E/E$_{ankle}$)$^{-0.5}$)
over the Galactic Gamma-Ray Burster (GGRB) source spectrum
(thick line, a power law with differential power law index
of -2.2 below the knee and $\simeq$ -2.5 above it)
by way of trapping in the
Galactic halo magnetic fields. The ``GGRBs'' are identified with
bipolar SNe in the present paper.}
\label{figuspec}
\end{figure}
\section{Observational evidence in favour of CR sources located
mainly (but nor completely) outside the solar circle}
\label{clue.l}
\subsection{Galacto-centric gradient of cosmic rays
from $\gamma$-ray data}
\label{gammag}
The analysis of EGRET data from Mattox and Strong\cite{mattox96}
and Erlykin et al.\cite{erlykin96}
find that the cosmic-ray 
density is constant within about 25 $\%$ between 5 and
15 kpc (5 - 18 kpc in Erlykin et al.\cite{erlykin96}) 
distance from the Galactic centre.
In this range the SNR- and pulsar-densities density decrease
by about a factor 5 to 10. 
Were these object classes the main sources of CRs
- and taking into account
CR diffusion in a large halo -
a much larger decrease
of CR density by a factor $\ge$ 3 over this range of 
Galacto-centric distances
would be expected\cite{bloemen93}.
The last bin (15 to 30 kpc in Mattox and Strong\cite{mattox96}) shows
a density about 30 $\%$ lower than the previous ones.
The analysis of the EGRET team\cite{huntero97} shows a 
drop in the density by
about 30 $\%$ in the range 5 - 10 kpc than a {\it rise} in density
up to 15 kpc in 3 quadrants.
At larger Galacto-centric distances 
a precipitous drop in density was inferred.
\\
These results are not in complete agreement with each other
but they
{\it do suggest that the universal sources of Galactic CRs are 
{\bf mainly}
located in shell at about 20 kpc distance from the Galactic
centre.} In this approximation one expects a constant 
density at smaller
distances and a rapid drop at larger distances; this is 
in agreement with the data. 

\subsection{Anisotropy of arrival directions of Galactic cosmic rays}
\label{anisec}
\begin{figure}[ht]
\vspace{0cm}
\hspace{0cm} \epsfxsize=9.1cm 
\epsfbox{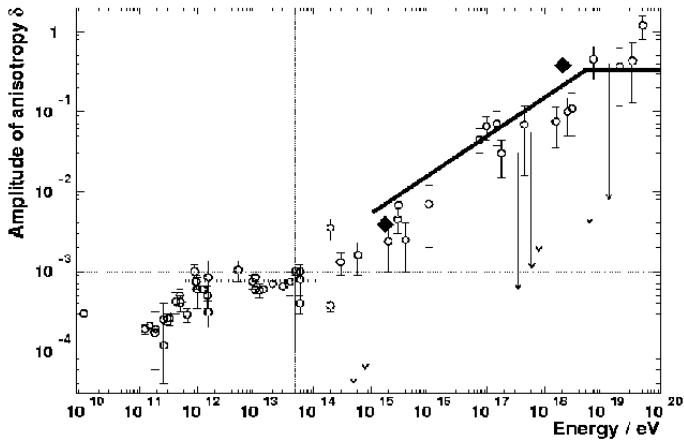}
\vspace{0cm}
\caption{World-data set as of 1997 of 
measurements the anisotropy amplitude of the first harmonic 
as a function
of energy. 
From the thesis of Schmele\cite{schmele}
where all references are given. The thin dotted lines 
indicate the data point obtained in this thesis from
data obtained by the HEGRA air-shower array.
The diamonds indicate the recent
results by Clay et al.\cite{clay97} (near 1 PeV) and 
Hayashida et al.\cite{hayashida98} 
(near 1 EeV), discussed 
in the  text. The quoted errors of these two determinations are smaller
than the symbol sizes.
The thick, full line indicate the anisotropy
theoretically predicted (eq.(\ref{ani}) with $f$=0.15 
(section \ref{ani_bel_ank})).
The thick dotted line indicates the constant anisotropy at energies
between about 1-100 TeV probably 
due to an anomalous motion of the solar system
relative to the rest frame of CRs (see text).
}
\label{figaniso}
\end{figure}

\begin{figure}[ht]
\vspace{0cm}
\hspace{0cm} \epsfxsize=9.1cm 
\epsfbox{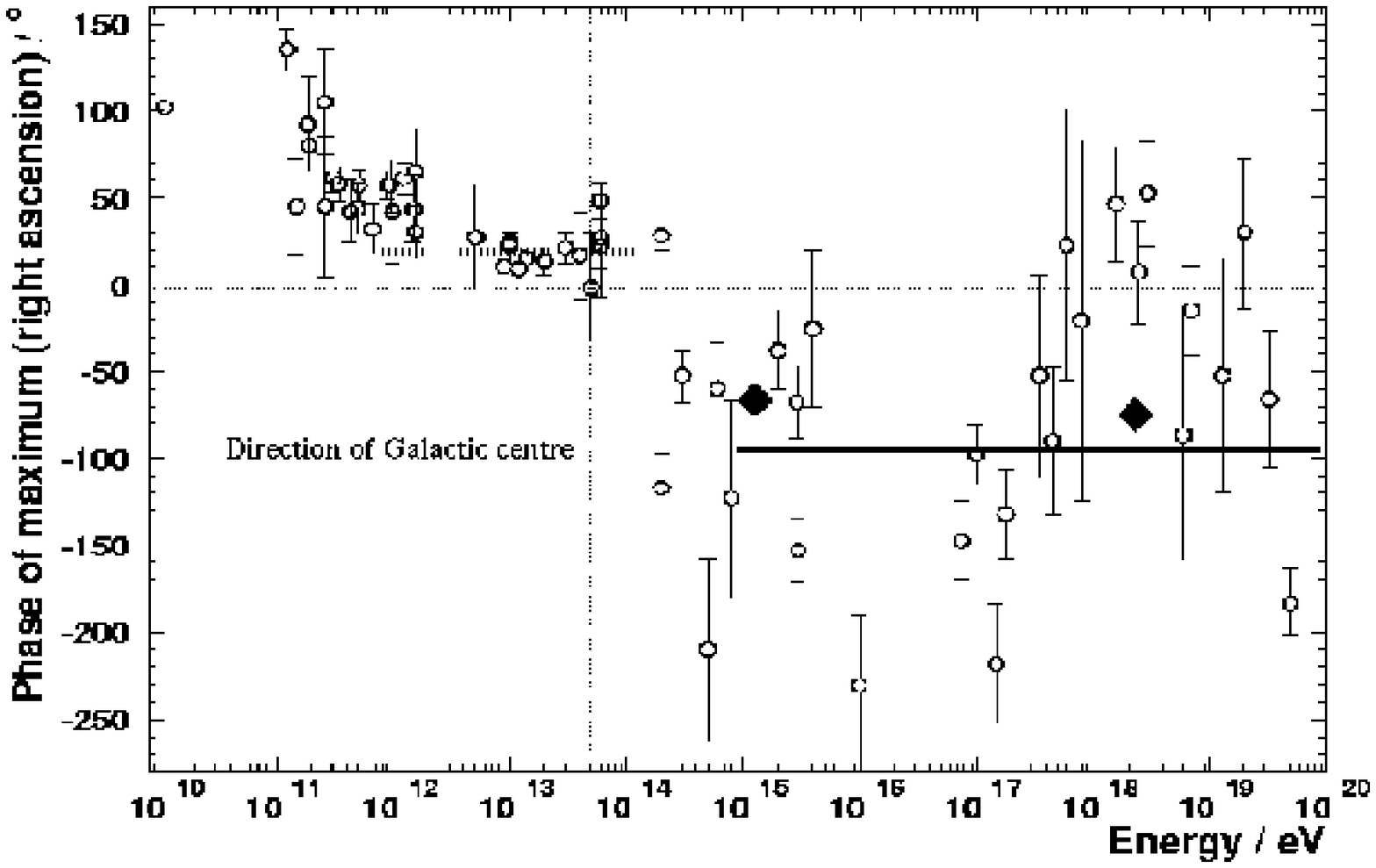}
\vspace{0cm}
\caption{World-data set as of 1997 of 
measurements the direction of the maximum
of the CR intensity in right ascension
as a function
of energy (first harmonic). 
From the thesis of Schmele\cite{schmele}
where all references are given. The thin dotted lines 
indicate the data point obtained in this thesis from
data obtained from the HEGRA air-shower array.
The thick, full line indicates the 
direction towards the Galactic centre.
The diamonds indicate the recent
results by Clay et al.\cite{clay97} (near 1 PeV) 
and Hayashida et al.\cite{hayashida98}
(near 1 EeV) discussed 
in the  text. The error bars of the result of Clay
et al. (1998) are comparable to the symbol size, no error
was given in Hayashida et al.\cite{hayashida98} 
for the determination of the direction.
The thick dotted line indicates the range where an energy-independent
anisotropy was found (see text).
}
\label{figaniso2}
\end{figure}

If {\bf all} universal CR sources were located outside the solar
circle one would
expect no anisotropy of Galactic cosmic rays at any energy:
there is none inside a shell isotropically 
emitting cosmic rays.
There is however experimental evidence for small anisotropies 
in Galactic cosmic rays at all energies\cite{linsley83}.
The observationally well established 
energy-independent anisotropy with an 
amplitude of $\approx$ 8 $\times$ 10$^{-4}$ at energies
between about 1 and 100 TeV is probably the result
of a motion of the local rest frame of cosmic
rays relative to the solar system with a speed of about 80 km/sec
via the Compton-Getting effect\cite{comget}.
The nature of this motion is unclear. Schmele\cite{schmele} recently proposed
an anomalous velocity component
of the solar-system surroundings
connected with a motion of the
``local fluff'' out of the centre of the
local super-bubble in which the CRs are at rest.
If this interesting suggestion is correct,
the anisotropy below  10$^{14}$ eV
contains information about the plasma that confines CRs
but holds no direct clues about the
origin of cosmic rays. 
\\
At higher energies   
there is a large world-data set which indicates an
anisotropy that increases with energy E as
E$^{0.5}$ up to the highest energies (fig. \ref{figaniso}) 
\cite{linsley83}.
In spite of a very large systematic 
scatter, the measurements of the direction of first-harmonic maximum
intensity in the total
world-data set seem to cluster around the Galactic-centre direction 
(see fig.\ref{figaniso2}).
Because individual anisotropy measurements usually have marginal statistical
significances and the derived 
locations of maximum intensity do not agree well,
doubts have been expressed if these results are more than upper limits
on a possible anisotropy.
However, a recent measurements and
reanalysis indicate significances for anisotropies
above 5 $\sigma$. 
\\
Clay et al.\cite{clay97} reanalysed the world data set
in the southern hemisphere
in the energy range 10$^{15}$ - 3 $\cdot$ 10$^{15}$ eV and
found an anisotropy amplitude of 
\begin{equation}
\delta_a = 0.33 \pm 0.06 \% 
\label{anisouth}
\end{equation}
which corresponds to a 
5.5 $\sigma$ effect. 
The direction of maximum intensity has a right ascension of
19.7 $\pm$ 0.7 h, not far from the one of the Galactic centre
(at 17.8 h). 
Hayashida et al.\cite{hayashida98} report an anisotropy 
at energies above 10$^{18}$ eV of 3.9 $\pm$ 1$\%$
with a statistical significance of 5.4 $\sigma$. 
Again an excess mainly from the Galactic-centre direction is indicated.
These two high-significance results are in reasonable agreement
with each other if the
E$^{0.5}$ energy dependence of the anisotropy amplitude
of the world data set is assumed. 
A more detailed future analysis needs to take into account
the different geographical locations of the various experiments which
necessitate small corrections of the measured anisotropy amplitudes
and directions.
\\
Summarising, two new high significance measurements confirm
the long known indication of an anisotropy amplitude rising
with square root of energy and prefer an origin of this anisotropy
of arrival directions from the direction of the Galactic centre.
\\
A natural origin for the observed anisotropy
is a net outward-streaming motion of cosmic rays produced in the
Galactic centre region. 
If such a streaming motion exists, the Compton-Getting
effect\cite{comget} leads to a large scale anisotropy with a maximal
intensity near the Galactic centre.
Let us estimate the expected magnitude an 
anisotropy $\delta_a$ produced in this way.
The preferred values of the
energy dependence of the diffusion constant D, halos size h
and spectral index $\gamma$ as 
discussed in sections \ref{propa} and \ref{ghalo} were chosen.
Here ``$f$'' is the fraction of
CR sources that are located within the solar circle.
\begin{eqnarray}
\delta_a = 0.5 \% (D/9 \cdot 10^{28} cm^2/sec)/ 
(h/20 kpc)/c  \times
\nonumber
\\
((\gamma/2.7)+1) (f/0.1)
\cdot ((E/10^{15} eV)/ 40 GeV)^{0.5} 
\label{ani}
\end{eqnarray}
This expression is valid from about 10$^{15}$ eV (at lower energies
another source of anisotropy dominates) to about 3 $\times$ 10$^{18}$ eV
(where CRs become unconfined to the Galaxy).
Here ``$f$'' is the fraction of
CR sources that are located within the solar circle.
Eq. (\ref{ani}) is plotted in fig.\ref{figaniso} and is seen
to agree with the data within the rather large errors. 
\\
{\it A consistent explanation of the CR anisotropies at all energies
is thus possible if f $\approx$ 10 $\%$
of the universal sources are located in the Galactic
centre and 90 $\%$ outside the solar circle.}
This is consistent with the observational constraints of section \ref{ghalo}
and the prediction within the cannonball model (section \ref{ani_bel_ank}).

\section{A candidate for the ``universal'' CR sources: ultrarelativistic 
ejecta from bipolar supernovae (aka ``cannonballs'')}
\label{sn.l}

\subsection{Ultrarelativistic ejecta (``cannonballs'') from supernovae}
\label{cb}
\subsubsection{The cannonball model of Gamma-Ray Bursts}
\label{cannon}
Recently Dar and De R\'ujula\cite{darrujula,darrujula2} 
developed the ``cannonball
model'' for Gamma-Ray Bursts (GRBs). Similar to
Woosley's\cite{woosley93} ``failed supernova'' concept
it proposes that 
the mechanism for core-collapse supernovae can 
involve a phase where matter falls onto a central 
compact object from large distances.
Episodic accretion onto this central
black hole or neutron star leads to the subsequent 
repeated emission of 
distinct masses of plasma,the ``cannonballs'' (fig.\ref{fig4}), 
initially moving with a Lorentz $\Gamma$ $\approx$ 1000.
Radiation emitted by the cannonballs is quantitatively
shown to explain the properties of GRBs.
This radiation is emitted isotropically in the rest frame of
the cannonball, but in the observer frame the Lorentz boost
strongly collimates the radiation into a cone 
with opening angle $\theta = $1/$\Gamma$
along the direction of motion (``relativistic beaming'').
Therefore only a {\it very small} fraction $\approx$ $\theta^2$/4
$\approx$ 2 $\times$ 10$^{-7}$
of all cannonballs are visible.
\\
Recently I found\cite{plaga00} that it is possible to quantitatively
understand a ``cepheid-like'' relationship  between variability 
and total luminosity of GRBs\cite{fenimore2000}
and between the width and height of 
spikes within a given GRB\cite{feni2} within the cannonball model.
This makes it plausible that the broad features of the
cannonball model are correct.
I will assume that this is the case in the rest of this paper.
\\
The initial energy of these cannonballs is not well known.
If the core of a pre-collapse
star is rapidly rotating, $\approx$ 3 $\cdot$ 10$^{53}$ erg
may be stored as rotational energy of the collapsed object,
and the jet seems the most effective way to dispose off
this energy\cite{pac}. I will assume below a total energy 
of 10$^{53}$ erg ejected in each of the symmetric 
cannonball jets in a SN
(rather higher than what is normally assumed but in line with
the expectations of Dar and De R\'ujula\cite{darrujula2}).
Smaller values lead to conditions in which CRs cannot be accelerated
to the observed ultra-high energies (section \ref{syro}).

\subsubsection{The frequency of supernovae that eject cannonballs,
inferred from the observed GRB rate}
\label{rate}
For some GRBs
Fenimore and Ramirez-Ruiz\cite{fenimore2000} infer redshifts in excess of 12
from their variability - luminosity relation. Because 
it seems unplausible that much larger
redshifts occur, within the cannonball model this indicates that 
cannonballs ejected exactly into the observer direction are visible from
the entire universe.
If that is true, the rate of cannonball-ejecting supernovae  
in our Galaxy is given as\cite{dar99}:
\begin{eqnarray}
R_{\rm prog} = {1\over{50 {\rm years}}} {R_{\rm GRB}\over{10^3/{\rm year}}} \times 
\nonumber
\\
(1/\left(({\theta_{\rm mean}^2}/5) \over{4(\Gamma/1000)^2}\right)) \times
\nonumber
\\ 
((L_B({\rm Galaxy})/2.3 \times 10^{10} L_{\odot})/
\nonumber
\\
(\rho_B/1.8h 10{^8} L_{\odot} {\rm Mpc}^{-3}))/
\nonumber
\\
(R_{\rm SFR(z=0)}/(\int (1+z)^{-1} R_{\rm SFR(z)} (dV_c/dz)dz)) \times 15
\label{freq}
\end{eqnarray}
Here $R_{\rm GRB}$ is the observed total rate of GRBs\cite{fishman}, 
L$_B$ is the
B-luminosity of our Galaxy, $\rho_B$ is the B-luminosity density in the
local universe\cite{loveday}, 
R$_{\rm SFR}$ is the star formation rate at a respective
redshift derived from optical observations\cite{lilly}.
The preferred value was derived with
a volume element d$V_c$ for $\Omega$=1 and $\Lambda$=0. 
$\theta_{\rm mean}^2$ is the mean squared angle between an observed
GRB and the observer divided by $\Gamma^2$. 
This angle is expected to be somewhat larger
than 1/$\Gamma^2$ because nearby 
GRBs are bright enough for detection
also at viewing angles $>$ 1/$\Gamma$.  
The mean($\theta^2$) $\approx$ 5 
was inferred by calculating the 
$\theta^2$ for all bursts in Fenimore and Ramirez-Ruiz\cite{fenimore2000}
using relations valid in the cannonball model\cite{plaga00}. 
An isotropic
luminosity at $\theta=0$ of 2 $\times$ 10$^{55}$ erg/sec 
(the highest luminosity  of any GRB in their sample) 
was assumed for this estimate.
The preferred value of  $R_{\rm prog}$ =
1/50 years corresponds to the estimated
total rate of all core-collapse supernova in 
our Galaxy\cite{scalo}, thus
indicating that {\it all} core-collapse supernovae eject cannonballs.
\\
The major uncertainty in the estimate of $R_{\rm prog}$ is the
dependence on the star-formation rate (SFR) on z. In particular
Fenimore and Ramirez-Ruiz\cite{fenimore2000} derive a 
$R_{\rm SFR-GRB}$ based on observations of GRB space densities alone.
This SFR continues to rise at redshifts beyond 1, in contrast to results
based on optical observations.
Using these results I obtain
$R_{\rm SFR-GRB(z=0)}/
\int (1+z)^{-1} R_{\rm SFR-GRB(z)}(dV_c/dz)dz$ $\approx$ 250.
This corresponds to $R_{\rm prog}$ $\approx$ 1/800 years.
The possible choice of other world-models introduces another uncertainty
of about a factor 2 in this estimate, so the plausible range
of rates of cannonball-producing SNe is about 1/50 - 1/1600 years. 
Preferred candidates for jet-forming core-collapse 
events have similar formation rates: SN Ib/c supernovae (R
$\approx$ 1/300 years\cite{scalo}), 
``failed supernovae'' (R $\approx$ 1/1000 years\cite{woosley93}
) and ``collapsars''
(R $\approx$ 1/1500 years\cite{woosleyfreq}).
\\
Summarising, based on the observed GRB rate,
the progenitor rate for cannonball ejecting supernova
could lie between about 0.05 - 1 of total core-collapse SN rate.

\subsubsection{The total cannonball-energy converted into CRs}
\label{lumi}
The ``CR luminosity'' L$_{\rm cr}$ - the energy per time in form of 
hadronic CRs injected into the Galactic confinement volume -
is given as:
\begin{eqnarray}
L_{\rm cr} = 2 \times 10^{42} {\rm erg/sec}
(R_{\rm prog}/1000 {\rm years}) \times
\nonumber
\\
(E_{\rm tot}/(
2 \times 10^{53} {\rm erg})) (\epsilon / 0.33)
\label{crlum}
\end{eqnarray}
Here $\epsilon$ is the conversion efficiency of total energy
to hadronic CRs within the {\it collimated} (section \ref{colli})
cannonball,
that is likely near the equipartition value of
1/3 in our scenario (see section \ref{syro}).
$R_{\rm prog}$ is the rate of cannonball-ejecting supernovae in
our Galaxy.
If cannonballs are the dominating source of hadronic CRs they
must supply the 
observed ``CR luminosity'' L$_{\rm cr}({\rm exp})$ $\approx$ 10$^{41}$ erg/sec
which can be inferred from experimental data\cite{bere}.
Within the plausible range for
$R_{\rm prog}$ discussed in the previous section,
less than 1/40 of the total cannonball-energy has
to be converted the energy of CR particles to explain the
observed CR luminosity L$_{\rm cr}$ (eq.(\ref{crlum})).

\subsection{Collimation of the cannonball}
\label{colli}
\subsubsection{Assumption of cannonball confinement}
\label{confine}
In the model of Dar and De R\'ujula\cite{darrujula}
the cannonball initially expands with the speed
of sound in a relativistic gas c/$\sqrt{3}$.
{\it It is a basic hypothesis of the present paper that some
collimation mechnism prevents any further expansion 
of the cannonball once
the cannonball has reached a limiting radius
R$_{\rm limit}$ $\approx$ 0.3 pc.}
If there is equipartition between magnetic-field energy
density, turbulent plasma motion and cosmic-ray energy density,
a total magnetic energy of E$_{\rm tot}$/3 $\approx$
3 $\times $10$^{52}$ erg (see sect. \ref{cannon}) corresponds to a magnetic
field inside the cannonball of 
B = $\sqrt{E_{\rm tot}}$/R$_{\rm limit}^3$  
$\approx$ 0.6 G.
\\
Efficient collimation is commonly observed
in extragalactic jets (for a review see\cite{narl}). 
There is no complete
theoretical understanding of the mechanisms leading to
collimation in general. Longair recently wrote\cite{longair97}:
``By which means relativistic jets are
collimated, is one of the major unsolved
problems of astrophsics. Almost certainly, the mechanism 
will require the presence of magnetic fields...''. 
In principle a slight deviation from equipartition
with strong magnetic fields might serve to contain
turbulent plasma and CRs.
However,
the ``plasma-virial  theorem'' forbids a purely 
magnetic stress balance
without additional outer stresses\cite{langmuir}. 
The only viable candidate for the outer stress in the
case of cannonballs seems to be ram pressure from the ambient
interhalo medium. It has been shown long ago 
that ram pressure can confine a spherical mass of plasma\cite{deyoung},
and Heinz recently argued that the same mechanism might lead
to confinement of ``bullets'' ejected by GRBs\cite{heinzthesis}.
Equating the typical expected ram pressure onto a subrelativistic
CB\footnote{see discussion in section \ref{syro} of the CB
propagation} P${_{\rm ram}}$
= m$_p$ (n$_{\rm interhalo}$/(10$^{-3}$/cm$^{-3}$)) ($\Gamma$/1.2) c$^2$
with its internal magnetic-field pressure B$_{\rm int}^2$/4$\pi$
shows that with the parameters indicated above
ram pressure can only balance an internal
field B$_{\rm int}$ of about 10 mG, about a factor 50 smaller than required.
Efficient collimation therefore requires a magnetic field that increases
towards the interior of the CB (similar to what happens in a
tokamak). 
What magnetic-field configuration of the cannonball would allow
a stable confined configuration in conjunction with external ram pressure
is an open question, closely related to Longair's ``open problem'' mentioned
above. As a detailed simulation of the internal dynamics is beyond the
scope of the manuscript, 
below I make the simplest possible assumption of free expansion
until a limiting radius R$_{\rm limit}$ is reached.
When the cannonball has slowed down to nonrelativistic
speeds and the ram-pressure becomes very small, a final
Sedov-Taylor expansion must take place\cite{darrujula}.

\subsection{Motion of confined plasmoids in the Galaxy}

\subsubsection{The interstellar medium ambient to cannonballs}
\label{ambient}
Lingenfelter, Higdon and Ramaty\cite{lingen} have pointed out that 
about 85 $\%$ of all core-collapse supernovae are expected
to occur in ``super-bubbles''. These are cavities in the interstellar
medium (ISM) blown by repeated SN explosions and filled with
hot (10$^{6}$ K) and tenuous (number density 10$^{-3}/{\rm cm}^3$) gas.
Their conclusion seems compelling to me. It is
an undisputed observational fact that about 50 $\%$
of the ISM volume of the Galactic disk is filled with super bubbles
\cite{spitzer}. Therefore at least half of all supernovae are expected 
to occur in super bubbles.
The actual fraction must be somewhat higher than this because
core-collapse supernovae are known to be spatially and temporally
correlated.
\\
The typical super-bubble size is several hundred parsecs, and
a large fraction of the super-bubble borders on the Galactic-halo
medium. The Galactic halo is observationally known to contain
hot, tenuous plasma which has similar properties (and possibly the same
origin) as super-bubble medium.
At a distance of 50 kpc from the Galactic centre 
the halo density was roughly determined
as 10$^{-4}$/cm$^{-3}$\cite{weiner} and 
from absorption studies the halos of spiral
Galaxies are known to extend to several hundred kpc\cite{lanzetta}.
\\
To good approximation the cannonballs are thus expected to propagate
in 85 $\%$ of all cases in super-bubble and halo medium and
in 15 $\%$ of all cases in the normal interstellar medium (with a density
of $\approx$ 0.1 cm$^{-3}$).
In addition, in
the former case denser material ($\rho$ $\approx$ 0.1 cm$^{-3}$)
from the progenitor star of the
SN in about the first parsec of propagation is expected.

\subsubsection{Cannonball propagation in interstellar medium}
\label{cannonprop}
When a mass of magnetised plasma (plasmoid) 
moves through the ionised interstellar 
medium two limiting cases can be distinguished.
\\
1. If the plasmoid is impenetrable to the incoming charged
particles (e.g. because of a homogenous, strong B field),
so that all are effectively reflected,
a collisionless bow-shock forms and the particles flow
around the plasmoid without any sweep up of ambient matter.
\\
2. If the incoming charged particles can freely enter the plasmoid
(e.g.because of a very inhomogeneous B field) no shock forms and
all incoming particles are swept up.
\\
Because cannonball B-field is expected to be strong but very
turbulent I think an intermediate situation is most likely.
The moving cannonball drives an ultrarelativistic collisionless shock
into the interstellar medium, but most of the incoming particles
(including the ones accelerated at this shock)
are eventually swept up by the following cannonball.

\subsubsection{Numerical calculation of cannonball motion}
\label{motion}
\begin{figure}[ht]
\vspace{0cm}
\hspace{0cm} \epsfxsize=7.1cm 
\epsfbox{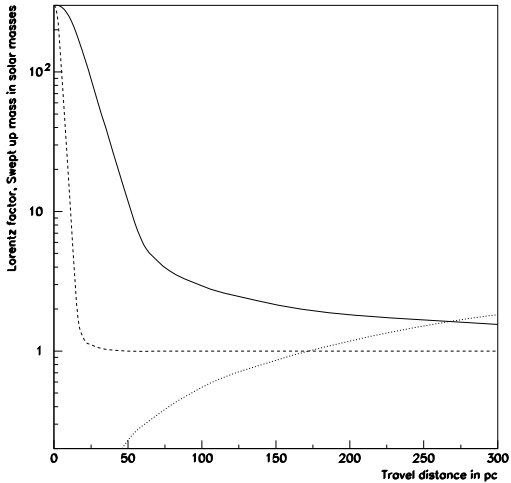}
\vspace{0cm}
\caption{
The evolution of the Lorentz factor of a plasmoid 
with the properties described in the text as function of
distance travelled in the observer frame. The full curve is
for motion in a tenuous super-bubble medium (case 1), the dashed
one for the normal warm interstellar medium (case 2). The dotted line
shows the amount swept-up matter in the latter case in units
of solar masses for case 2.
}
\label{fig1}
\end{figure}

\begin{figure}[ht]
\vspace{1cm}
\hspace{0cm} \epsfxsize=7.1cm \epsfbox{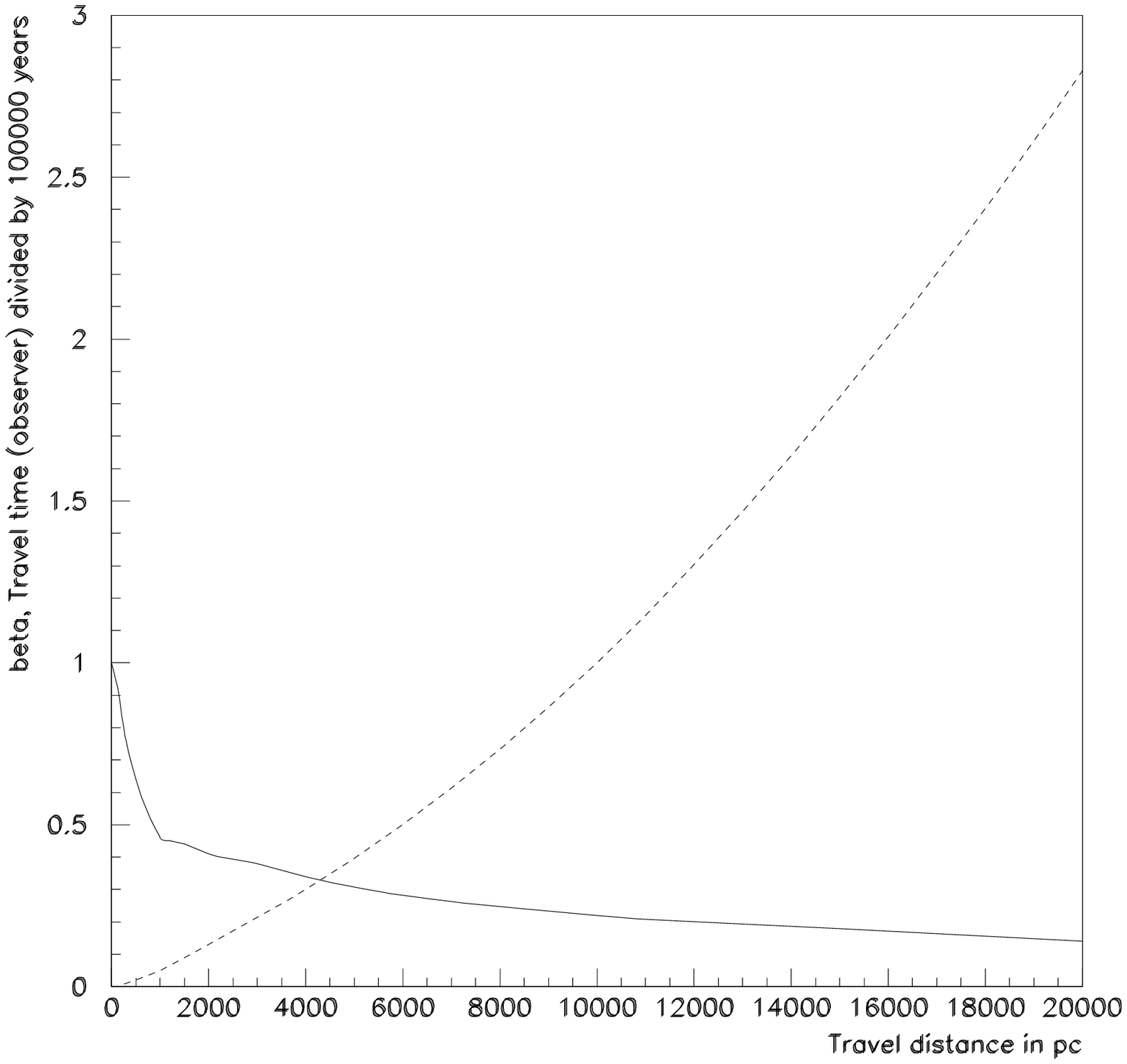}
\vspace{1cm}
\caption{
The $\beta$=v/c factor of a plasmoid propagating in a super-bubble
environment (case 1)
as a function of transversed distance in the observer frame.
The dashed curve gives the elapsed time in the observer frame.
}
\label{fig2}
\end{figure}

The following relativistic equations of motion 
for plasmoids ejected with an initial Lorentz factor of
$\Gamma$ $\approx$ 300 and an initial energy of E $\approx$ 10$^{53}$ ergs
were numerically integrated using the following expressions
\cite{dermer2}:
\begin{eqnarray}
d \Gamma/dm = - (\Gamma^2-1)/M
\\
dm/dr = \pi R^2 \rho(r)
\\
dM/dr = \Gamma dm/dr
\end{eqnarray}
m is swept up mass, $\rho$ the ambient density 
 and r the distance travelled by the cannonball.
M is the cannonball mass, and was set to E/$\Gamma$
initially.
The cannonball radius 
R was calculated assuming that the plasmoid expands with 
a speed of v = 0.2 c/$\Gamma$ in the observer frame
until it reaches a radius of 
R$_{\rm final}$ $\approx$ 0.3 pc.
 No further expansion 
was assumed to take place afterwards.
v is similar to the initial expansion in the plasma clouds ejected
by GRS 1915+105 and also the speed of sound
in a relativistic gas.
\\
As discussed in section \ref{cannonprop}
the plasmoid is
assumed to sweep up all ambient matter, i.e. it grows in mass.
The calculation does not include a treatment of the internal
dynamics of the plasmoid, the matter is assumed to move with the
plasmoid´s Lorentz factor at the time of sweep up.
Therefore the presented results are only indicative
and unrealistic, especially in the final stages of propagation
when most of the initial kinetic energy has been transformed into
turbulent plasma motion, B fields and CRs.
\\
For the ambient matter the two cases discussed in 
section \ref{ambient} were assumed.
In {\bf case 1} the cannonball-plasmoid 
first moves in matter of 0.1 cm$^{-3}$
for 1 pc (stellar wind), 1 kpc in 10$^{-3}$ cm$^{-3}$ (super bubble)
and then in 2 $\cdot$ 10$^{-4}$ cm$^{-4}$ (halo).
In {\bf case 2} the plasmoid moves in matter with a density
of 0.1 cm$^{-3}$ throughout.
\\
Some relevant parameters as a function of plasmoid travel distance
are given in figs.\ref{fig1} and \ref{fig2}. 
\\
{\it Case 1 ($\approx$ 85 $\%$ of all cannonballs, see section \ref{ambient})}
\\
The
plasmoid is slowed down non-ultra-relativistic speeds ($\Gamma$ $<$ 10)
in the first ca. 40 pc. 
It then travels for a time t$_p$ 
(some hundreds of thousands years)
with subrelativistic speed ($\beta$ $\approx$ 0.2). 
The number of cosmic-ray producing plasmoids 
at any time is given as t$_p$/t$_i$ where t$_i$ is the mean interval
of plasmoid producing events (estimated as 50 - 1600 years in section
\ref{rate}). I thus expect some hundreds to thousands of 
plasmoids in the halo that are still actively 
accelerating and releasing
CRs.
\\
{\it Case 2 ($\approx$ 15 $\%$ of all cannonballs, see section \ref{ambient})}
\\
The cannonball Lorentzfactor $\Gamma$ falls below 10
at about 10 pc and its speed $\beta$ below 0.1 at 80 pc.
The cannonball does not travel far from its point of origin 
in a SN, i.e. it remains close to the Galactic centre in
general. Consequently also all CRs produced by the cannonball
will be released there, within the solar circle.
\\
These results were found to be very insensitive on the assumed
initial value of $\Gamma$. However, if the initial energy E is
assumed to be significantly smaller or the final confined plasmoid
radius R$_{\rm limit}$ significantly larger, all cannonballs range
out before reaching the Galactic halo.

\section{Acceleration of cosmic rays by and in cannonballs}
\label{accel}
\subsection{1st-order Fermi acceleration up to the knee: ultrarelativistic
shocks}
\label{1st}
Heavens $\&$ Drury\cite{heavens} - in a pioneering paper on acceleration
in relativistic shocks -
made the conjecture that the 
power-law index of particles accelerated
in such shocks asymptotically approaches 
a universal index of $\delta$ = $-$2.2 (eq.(\ref{source2}))
with rising Lorentz factor. This 
has recently been confirmed by several researchers
\cite{kirk,bednarz,achterberg}.
As this is exactly the source index of the ``universal CR sources''
in section \ref{sourcespec}, I propose that the bulk of Galactic hadronic CRs
is accelerated at the ultra-relativistic shocks driven by cannonballs
into the interstellar medium.
Extragalactic jets, which might be qualitatively 
similar objects, have been identified as prolific CR
accelerators\cite{burb}, in particular their hot spots\cite{bira}.
\\
The accelerated particles are immediately swept up into the plasmoid
(section \ref{cannonprop})
and remain confined there until it has slowed down to subrelativisitic
speeds in the Galactic halo (section \ref{syro}).
In this scenario a spectrum extending down
to small energies with the correct index is expected.
The release mainly outside the solar circle leads to a
spatial ``source distribution'' in agreement with the one
expected for ``universal sources'' (section \ref{ghalo}). 
\\
The maximal energy that can be reached in this acceleration process
is limited by the extension of the shock
and the {\it interstellar} magnetic field (the shock propagates
in this medium) and is given as\cite{achterberg}:
\begin{equation}
E_{\rm max} = 3 \cdot 10^{15} eV (R/0.02 {\rm pc}) (Z B/\mu G) (\Gamma / 50)
\label{knee}
\end{equation}
R is radius of the plasmoid when it has a Lorentz factor
of $\Gamma$ = 50. The preferred value of R=0.02 pc was chosen based
on results of the numerical calculation discussed in
section \ref{motion}.
Z is the charge
of an accelerated particle.
A precise calculation of this maximum energy would require
a more detailed calculation of the plasmoid dynamics than I
have performed, but the inserted numbers are correct to the
order of magnitude. I identify E$_{\rm max}$
with the ``knee energy'' E$_{\rm knee}$ (sect \ref{sourcespec}).
The observed knee seems to be a rather ``sharp'' feature in the
primary CR spectrum\cite{arqueros}.
In my scenario there are $\approx$ 1000 current sources for these CRs.
It seems highly unlikely that they would be so uniform
in their properties that a sharp knee would be a natural
feature\footnote{I thank an anonymous referee for pointing this out
to me}. Therefore the sharpness of the knee remains unexplained.

\subsection{2nd-order Fermi acceleration above the knee: 
the Syrovatskii mechanism}
\label{syro}
 
While the plasmoids are ``coasting''
in the super-bubble (case 1 of section \ref{motion}) or 
normal interstellar medium (case 2)
for  a time t$_{\rm coast}$ $\approx$ several 10$^5$ years (case 1) resp.
hundreds years (case 2)
with ``subrelativistic'' speeds
\footnote{defined here as 0.2 $<$ $\beta$ $<$ 0.5)} 
internal turbulent plasma motions will further accelerate
the hadronic ``swept up'' CR pool via the second-order Fermi mechanism.
The coasting time scales t$_{\rm coast}$
for case 1 and 2 were estimated from the
numerical calculation explained section \ref{motion} (see also fig.\ref{fig2}).
\\
Without turbulent plasma motion the swept-up CRs remain very
efficiently confined in the plasmoid during the coasting phase
except at the highest, near-knee energies.
The diffusive-escape time scale from a sphere with radius R
and a mean free scattering path $\lambda$
is given as  t$_d$=1.5 R$^2$/c$\lambda$ \cite{hillas84}.
Here $\lambda$ is a particle´s mean free path for scattering
from magnetic inhomogeneities. I set $\lambda$ to its Larmor radius
in this manuscript always (i.e. within the cannonball
I always assume the Bohm limit).
One then gets for the escape time from the plasmoid with
the ``confined'' radius R$_{\rm limit}$ 
and internal magnetic field strength B (section \ref{confine}):
\begin{eqnarray}
t_d=  10^5 
(R_{\rm limit}/0.3 pc)^2 \times 
\nonumber
\\
(B/0.6 G)/(E_i/3 \cdot 10^{15} eV)
{\rm years} \approx t_{\rm coast}
\end{eqnarray}
At energies below E$_{\rm knee}$ t$_d$ exceeds t$_{\rm coast}$
by a large margin.
The rate of diffusive release - which determines the total
CR-luminosity (section \ref{lumi}) -
is then determined by 
the turbulent plasma motion in the plasmoid, and I do
not try to determine it here.
\\
It is plausible that equipartition between turbulent motion,
magnetic fields and cosmic rays holds in the confined plasmoid.
The major energy loss then occurs via loss
of cosmic-rays due to diffusion or release in bunches due to disturbances
of the magnetic field at the plasmoid boundary.
Syrovatskii\cite{syrovatsky61} has shown
that in this situation equipartition is restored through
particle acceleration and the released CR particles are distributed
in energy according to 
a power-law spectrum with a universal differential index.
In particular he shows:
\\
At energies that were not previously populated in the initial
pool of CRs -  i.e. in our case above the ``knee'' energy - 
the universal differential power-law index of the released CRs
according to eq.(\ref{source2})
is $\delta_2$ = $-$2.5.
\\
Moreover, it can be easily shown with the expressions
in the appendix of  Syrovatskii's paper \cite{syrovatsky61}:
\\
At energies which were populated in the initial
pool of CRs with an index of $\delta_1$=$-$2.2 - i.e. in our case
below the ``knee energy'' - the universal differential index
of the released CRs is (again) $-$2.2.
This ``invariance'' only holds for the index of $-$2.2 and is
thus apparently fortuitous.
\\
Syrovatskii's\cite{syrovatsky61} results are independent of any details
of the Fermi acceleration process because they follow from
purely thermodynamic considerations.
\\
I thus obtain the source spectrum that we had derived as most
likely for
the ``universal source'' class in section \ref{sourcespec},
a power law with an index of $-$2.2 below the knee and $-$2.5 above
the knee up to the highest observed energies
{\it I therefore propose that hadronic CRs mainly stem from
cannonballs in the Galactic halo that release previously
accelerated particles.}
\subsubsection{Possible visibility of 
Galactic cannonballs in the radio range}
The expected radio luminosity of
cannonballs in our Galaxy
is rather smaller than one might expect at first sight.
In their early ultrarelativistic stage 
relativistic beaming
prevents the visibility of Galactic cannonballs.
The emitted radiation is then visible only in external distant galaxies
as a GRB afterglow.
Galactic cannonballs become in principle 
visible only in the ``coasting phase'',
when no important beaming takes place and the
cannonball accelerates CRs via the Syrovatskii mechanism in the present
scenario.
In this phase nearly all of the CR acceleration has already
taken place in the present scenario. 
Only a small fraction  $f_{\rm 2nd-order}$ (see below)
of the total
CR energy produced by a cannonball is accelerated and can contribute
to the radio luminosity. 
\\
A fundamental uncertainty in predicting the cannonball radio luminosity
is the estimation of a 
factor f$_{\rm e/p}$, giving
the CR energy injected
into electrons relative to the one injected into nuclei in the coasting phase.
Because of the high magnetic field in the cannonball,
electrons have a very short
lifetime in the plasmoid due to synchrotron losses and need to
be injected during the ``coasting phase''.
It is a fundamental assumption of the Syrovatskii mechanism 
that the injection of new particles from the thermal
pool has ceased, so that mainly already-relativistic particles are
accelerated \cite{syrovatsky61}.
In this limit f$_{\rm e/p}$=0 and the radio-luminosity
of cannonballs is small. However, it seems likely
that this assumption holds only approximately and that f$_{\rm e/p}$ is 
smaller than 1 by some not extremely large factor.
The radio luminosity of a cannonball in the Galactic halo
at 1.4 Ghz $\phi_{\rm 1.4GHz}$ assuming a 
differential spectral index of the accelrated
particles of -2.5
can be parameterized as follows:
\begin{eqnarray}
\phi_{\rm 1.4GHz} \approx 3 (E_{\rm tot}/10^{53} {\rm erg}) \times
\nonumber
\\
(R_{\rm prog} / 1000 {\rm years}) \times 
\nonumber
\\
(f_{\rm 2nd-order} / 3 \cdot 10^{-5}) \times
\nonumber
\\
(t_{\rm coast} / 3 \cdot 10^5 {\rm years}) \times
\nonumber
\\
(20 {\rm kpc} / d_{\rm CB})^2 f_{\rm e/p} {\rm Jansky} 
\label{ralum}
\end{eqnarray}
Here $f_{\rm 2nd-order}$ is the ratio of the energy injected into
CRs in the second-order process relative to the total. The estimated
value corresponds to the energy content of CRs above the knee relative
to the total because in the present scenario CRs 
are reaccelerated to energies exceeding the ``knee'' energy
in the ``coasting phase'' of the cannonballs.
d$_{\rm CB}$ is the distance of the cannonball.
\\
With a factor f$_{\rm e/p}$ somewhat below 1 the expected radio brightness
of cannonballs is in the sub-Jansky regime. A systematic search for 
nonthermal radio-sources with a high proper 
motion (on the order of 1 arcsecond/year) is a crucial test of
the present scenario. 
There are 771000 radio sources the recent
FIRST catalogue \cite{first} of which only 19 $\%$ have optical counterparts. 
No systematic search
for high proper motions was performed in this data base.
If one embarks on one, there are many 
possible technical reasons for false positives\cite{white}.
Even at a flux level of one Jansky not all radio sources in the
FIRST catalogue are identified.
Radio sources with flux levels near 
1 Jy are frequently very faint high-z radio galaxies.
\\
In the absence of a proper-motion survey in
the radio and/or systematic identifications of all radio sources 
{\it with a limiting flux level below about
a Jansky}
the absence of ``obvious'' radio counterparts to cannonballs
cannot yet be an argument to reject the present scenario ``a limite''.
\\
When the cannonball finally stops, it enters
Sedov-Taylor phase. In this phase adiabatic losses
probably prevent an important release of CRs and its brightness
is difficult to predict.
For an earlier 
very preliminary discussion about expected $\gamma$-ray fluxes 
of cannonballs see Ref.\cite{plagaokkie}. 

\subsubsection{Can second-order acceleration in the plasmoids
reach ultra-high energies?}
\label{uhe}
To reach the highest observed CR energies  
UH energies (i.e. energy of protons up
at least E$_f$ $\approx$ 10$^{20}$ eV)
in second-order Fermi acceleration 
three conditions have to be fulfilled \cite{hillas84}.
\\
1. The lifetime of the plasmoid 
``$t_{\rm coast}$'' (see beginning of section \ref{syro})
has to exceed the time ``t$_n$''
necessary
for acceleration from the initial E$_i$
to the final energy E$_f$:
\begin{equation}
  t_{\rm coast} \geq {\rm hundreds\, of \, years} > t_n
\label{cond1}
\end{equation}
An analogous
condition limits the maximum energy that can be reached in
CR acceleration in SNR to about 100 TeV \cite{bere}.
t$_n$ is given as ln(E$_f$/E$_i$) $\times$ t$_a$.
The second-order Fermi acceleration time scale 
is t$_a$ = $\lambda$/2c$\beta^2$ \cite{hillas84}. 
For a conservative estimate of t$_n$ I choose the $\lambda$ for
the final maximum energy 
E$_f$ = 10$^{20}$ eV.
\begin{eqnarray}
t_n = 400 (\beta/0.5)^2 \times 
\nonumber
\\
ln((E_f/ 10^{20} eV) / (E_i/ 3 \cdot 10^{15} eV)) \times
\nonumber
\\
(E_f/10^{20} eV)/ (B/0.6 G)
{\rm years} 
\label{at1}
\end{eqnarray}
Taking into account that this estimate is conservative
I conclude that the condition (\ref{cond1}) can be surely met for protons
if $\beta$ $>$ 0.5.
\\
2. The diffusive-escape time scale t$_d$=1.5 R$^2$/c$\lambda$ 
must be larger than the second-order
acceleration time scale t$_a$=$\lambda$/2c$\beta^2$ 
up to the highest energies,
otherwise the particles escape before being accelerated.
I obtain for protons:
\begin{eqnarray}
t_d/t_a \approx (\beta/0.2)^2 (R_{\rm limit}/0.3 pc)^2 \times
\nonumber
\\ 
(B/0.6 G)^2 /(E/10^{20} eV)^2
\end{eqnarray}
The condition is thus fulfilled for $\beta$ $>$ 0.2.
\\
3. The synchrotron-loss time scale t$_s$ $\simeq$
k E$^{-1}$ B$^{-2}$  with k={${3 m^4 c^7}\over{2 e^4 sin^2(\alpha)}$,
has to remain larger
than the acceleration time scale t$_a$,
otherwise acceleration is shut off.
$\alpha$ is the pitch angle and is conservatively set to $\pi$/2 below.
\begin{eqnarray}
t_s = 3.9/((E/10^{20} eV) \cdot (B/0.6 G)^2) {\rm years}
\approx t_a 
\nonumber
\\
\approx 3.1 ({\rm E}/10^{20} {\rm eV})/(({\rm B}/0.6 {\rm G}) 
\times (\beta/0.3)^{2}) {\rm years}
\end{eqnarray}
The condition is seen to be fulfilled for $\beta >$ 0.3.
\\
I conclude that there is no basic argument
forbidding acceleration of UHE CRs in plasmoids.
However, it seems difficult to accelerate protons to
energies far above 10$^{20}$ eV with this mechanism.

\section{Cannonballs as universal CR sources and 
observational facts}
\label{ofacts}
\subsection{UHE CRs from plasmoids in the Galactic halo:
Greisen cutoff and anisotropies on a small angular scale}

If CRs originate in the Galactic halo, the
absence of the ``Greisen-Zatsepin cutoff''
- that is otherwise made quite puzzling by
the lack of extragalactic source candidates near the direction of
individual UHE-CRs \cite{takeda} -
is readily understood. At a distance of typically 20 kpc
no significant absorption of UHE protons 
in the 3 K$^\circ$ background radiation is expected.
\\
There is some evidence for a small-scale clustering in  the arrival
directions of UHE CRs \cite{cluster2}. The number of discrete
sources necessary to explain the observed degree of clustering
was estimated as $\approx$ 400\cite{cluster} in agreement with
the order of magnitude number of cosmic-ray producing plasmoids
estimated in section \ref{motion}.
Consequently
it is proposed to identify the subrelativisitic cannonballs 
in the Galactic halo as the sources responsible for the small-scale
clustering\cite{darplaga}.

\subsection{Expected anisotropy of CR
arrival directions from coasting cannonballs}

\subsubsection{At energies below the ankle}
\label{ani_bel_ank}
Core-collapse supernovae are well known to 
occur predominantly within the
solar circle \cite{bere}.
Therefore those cannonballs that range out in the dense disk-interstellar
medium (corresponding
to case 2 in section \ref{motion}) release all CRs at Galacto-centric
distances smaller than the solar one.
These CRs (estimated to be about 15 $\%$ of the total in section \ref{motion})
will contribute to the anisotropy of CRs at all energies.
We had seen in section \ref{anisec} that about this fraction 
$f$ (Eq.\ref{ani}) of CRs is required to be released
by the ``universal'' sources within the solar circle.
The cannonballs therefore remarkably ``fit the bill'' derived for
``universal sources''in section \ref{anisec}.

\subsubsection{At energies above the ankle}

The absence of significant large-scale anisotropies of
arrival directions
at UHE energies (section \ref{hlower}) is understood because
an observer inside an emitting sphere sees no anisotropies.
However about 15 $\%$ of all cannonballs are expected
to release CRs at Galactic locations similar to the ones
of SNe (section \ref{motion}).
Thus it is predicted that about 15 $\%$ of all CRs that
propagate roughly on straight lines in the Galactic magnetic field
(i.e. with energies well above the ankle) 
come from the general directions of the
Galactic centre and disk. Presently the world-data sample 
at these energies is not large enough to confirm or rule out this
prediction.

\subsection{Galacto-centric distribution of the CR density}

I had discussed in section \ref{gammag} that the very small observed
gradient of CR density as a function of Galacto-centric radius
strongly points towards CR sources mainly outside the solar circle.
The cannonballs are expected to release $\approx$ 15 $\%$ of all CRs
within the solar circle (section \ref{motion}), 
so a small non-vanishing gradient is expected. The data are in 
qualitative agreement with this expected ``cannonball density
distribution'' but the presence of large systematic errors seem to 
make a further detailed test of the scenario in this area
difficult.

\subsection{Predicted chemical composition of Cosmic Rays
at all energies}

One can now calculate the chemical composition of cosmic rays.
At low energies, below the knee, CRs are accelerated from the
local interstellar medium and are thus expected to 
have abundances at a given energy 
per nucleus similar to the solar ones, as observed.
I take the observed abundance at low energies as the
starting point.
In both first and second order Fermi acceleration
identical power-law indices for
all nuclei are expected.
The knee (eq. (\ref{knee})) 
and ankle positions in energy are proportional to Z.
With these assumptions 
the spectrum for all indices can be predicted  (fig\ref{chem1}).
We saw in section \ref{uhe}
that acceleration of protons
beyond $\times$ 10$^{20}$ eV is difficult in cannonballs
and therefore assume an exponential cutoff at an energy/nucleus above
Z $\cdot$ 10$^{20}$ eV. It is seen that in the universal
scenario the chemical composition gets heavier near the knee -
similar to what is expected
in many other models. Near the ankle the composition becomes
very light, similar to the expectations in
some extragalactic scenarios for the origin of UHE CRs
but finally, beyond
10$^{20}$ eV the composition becomes heavy again.

\begin{figure}[ht]
\vspace{0cm}
\hspace{0cm} \epsfxsize=9.1cm
\epsfbox{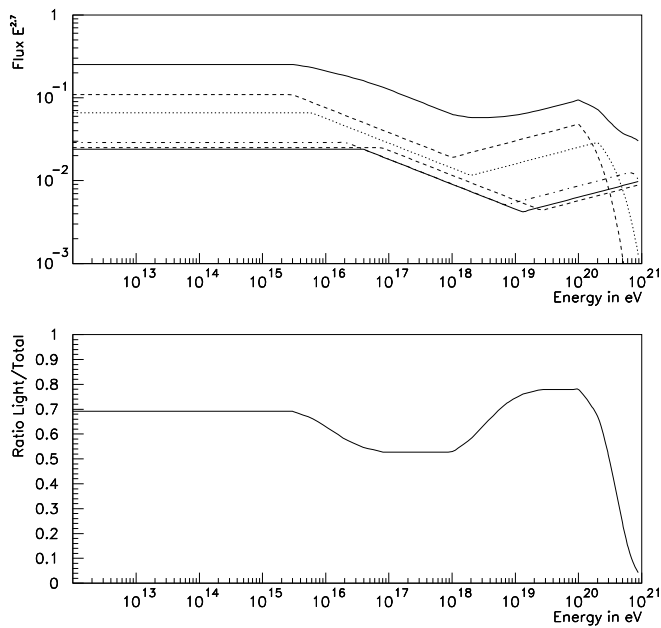}
\vspace{0cm}
\caption
{{\bf Upper panel.} The intensity of 5 chemical  groups
(in the order of falling intensity at low energies:
p (dashed),He (dotted),CNO group (dash-dot), Ne-S group
(dashed), Cl-Fe group(full)) 
in m$^{-2}$ s$^{-1}$  sr$^{-1}$ TeV$^{1.7}$
as a function of energy/nucleus.       
The uppermost line
is the sum of all components in the same units. 
The composition for low energies below the knee is taken from
Wiebel\cite{wiebel}. Above the knee the prediction of the present model
is shown.
{\bf Lower panel:} the ratio of the 
light element intensity (p + He) to the total intensity.
as derived from the upper panel.
}
\label{chem1}
\end{figure}

\section{Discussion}
\label{sum}
\subsection{Assumptions}
The {\it major} assumptions of the present model for CR origin are:
\\
1. The basic features of the ``cannonball model'' for GRBs
\cite{darrujula} are correct.
\\
2. The CR propagate diffusively in a relatively
large Galactic halo (radius $\approx$ 20 kpc). The energy
dependence of the diffusion coefficient is Kraichnanian at all
energies.
\\
3. The total energy of the ejected cannonballs is $\approx$ 10$^{53}$
ergs, about an order of magnitude higher than often assumed.
\\
4. After an initial expansion the cannonball remains confined
to a radius of about 0.3 pc, due to ram-pressure and magnetic effects
until it has slowed down to speeds below about 0.1 c.
\\
While there are theoretical and observational 
arguments in favour of assumption 1 and 2,
assumptions 3 and 4 have no better motivation presently than to make
the model work.
\\
Additional minor assumptions are discussed in the text.

\subsection{Achievements}

The location of CR sources at all energies mainly in the
Galactic halo and with a fraction
$f$ $\approx$ 15 \% near the Galactic centre
expected in our model
allows to understand the following
observations more naturally than many other models:
\\
1. The absence of the Greisen-Kuzmin cutoff 
together with the absence of a large scale
anisotropy at energies above about 4 $\times$ 10$^{19}$ eV and
extragalactic counterparts in the direction of the highest-energy
CRs.
\\
2. The observed anisotropy of CR with energies between about 100 TeV
and 10$^{18}$ eV quantitatively. 
\\
3. The Galacto-centric distribution of low-energy hadronic
CRs inferred from $\gamma$-ray astronomy.
\\
The origin of hadronic CRs in  
about 1000 cannonballs 
speeding through the interstellar and interhalo medium
allows to understand:
\\
4. The possible indication for small-scale clustering is
understood as being due to emission from individual cannonballs.
\\
5. The origin of the ``knee feature'' in the hadronic CR spectrum
and the form of the source-spectrum and higher and lower energies.
The observed ``sharpness'' of the knee remains unexplained, however.
\\
6. Finally the idea of a ``universal-source class'', accelerating 
CRs at all energies, leads to a very natural understanding of
the energy for the ``ankle'' feature in the spectrum of Galactic
CRs.

\subsection{Predictions}
The most characteristic prediction of a universal origin
of hadronic CRs at all energies is the variation of chemical
composition above the knee shown in fig.(\ref{chem1}).
Evidence for a Kolmogoroff rather than Kraichnan 
energy dependence of the diffusion coefficient 
for cosmic rays (e.g. from the ongoing 
experimental studies on reacceleration)
would rule out the ``universal scenario'' presented
here.
If ``cannonballs'' accelerate all CRs above the ankle,
on the order of 1000 spatially discrete sources of UHE CRs must be found
by the next generation of UHE-CR detectors.
About 15 $\%$ of these clusters must be preferentially spatially distributed
like core-collapse SNe (i.e. near the Galactic disk and centre).
\\
The counterparts of these sources must be very compact 
plasma clouds with subrelativistic proper-motions.
The present model stands or falls with the existence 
of counterparts to the cannonballs
in various regions of the electromagnetic
spectrum.

{\bf Acknowledgements}
I thank Maria Diaz for reading the manuscript and
Arnon Dar and Alvaro De R\'ujula for many helpful explanations
of the cannonball model. I am grateful to an anonymous referee for helpful
criticism.
The author is
supported by a Heisenberg fellowship of the DFG.}

\end{document}